\documentclass[runningheads]{llncs}
\def\doi#1{\href{https://doi.org/\detokenize{#1}}{\url{https://doi.org/\detokenize{#1}}}}

\usepackage{graphicx}
\usepackage{wrapfig}
\usepackage{soul}
\usepackage{color}
\usepackage{subcaption}
\usepackage{float}

\usepackage{listings}
\usepackage{diagbox}

\usepackage{xcolor}
\usepackage{textcomp}
\usepackage{paralist}

\definecolor{codegreen}{rgb}{0,0.6,0}
\definecolor{codegray}{rgb}{0.5,0.5,0.5}
\definecolor{codepurple}{HTML}{C42043}
\definecolor{codeblue}{HTML}{0000FF}
\definecolor{backcolour}{HTML}{FFFFFF}
\definecolor{bookColor}{cmyk}{0,0,0,0.90}  
\color{bookColor}

\lstdefinestyle{mystyle}{
	backgroundcolor=\color{backcolour},   
	commentstyle=\color{codegreen},
	keywordstyle=\color{codeblue},
	numberstyle=\numberstyle,
	stringstyle=\color{codepurple},
	basicstyle=\scriptsize\upshape\ttfamily,
	breakatwhitespace=true,
	breakautoindent=true,
	breaklines=true,
	breakindent=0pt,
	captionpos=t,
	keepspaces=true,
	numbers=left,
	numbersep=5pt,
	showspaces=false,
	showstringspaces=false,
	showtabs=false,
	tabsize=2,
	showlines=true,
    escapeinside=||,
	framesep=1pt,
	xleftmargin=18pt,
	framexleftmargin=18pt,
	frame=tb,
	framerule=1pt
}
\newcommand\numberstyle[1]{%
	\scriptsize	
	\color{codegray}%
	\ttfamily
	\ifnum#1<10 \fi#1|%
}

\newcommand{\event}{e}

\newcommand{\trace}{\tau}
\newcommand{\elog}{\mathcal{L}}

\usepackage{xcolor,colortbl}

\definecolor{Gray}{gray}{0.85} 
\definecolor{Grey}{HTML}{D6DCE4}
\definecolor{LightCyan}{rgb}{0.88,1,1}
\definecolor{grannysmithapple}{rgb}{0.66, 0.89, 0.63}
\definecolor{honeydew}{rgb}{0.94, 1.0, 0.94}
\definecolor{lavenderblush}{rgb}{1.0, 0.94, 0.96} 
\definecolor{lightblue}{rgb}{0.68, 0.85, 0.9}
\definecolor{lightapricot}{rgb}{0.99, 0.84, 0.69}

\newcolumntype{o}{>{\columncolor{Grey}}c}
\newcolumntype{x}{>{\columncolor{LightCyan}}c}

\usepackage{tikz}
\usetikzlibrary{decorations.pathreplacing}
\usetikzlibrary{calc}

\newcommand{\tikzmark}[1]{\tikz[remember picture,overlay]\node[yshift=1pt](#1){};}

\usepackage{array,multirow}

\begin{document}

\title{Efficient Checking of Timed Order Compliance Rules over Graph-encoded Event Logs} 

\author{Nesma M. Zaki\inst{1}\and Iman M. A. Helal\inst{1}\orcidID{0000-0001-8434-7551}\and Ahmed Awad\inst{2,1}\orcidID{0000-0003-1879-1026}\and Ehab E. Hassanein~\inst{1}}

\authorrunning{N. Zaki et al.}

\institute{Cairo University, Giza, Egypt \email{\{n.mostafa,i.helal,a.gaafar,e.ezat\}@fci-cu.edu.eg}\and University of Tartu, Tartu, Estonia}
\maketitle

\begin{abstract}
Validation of compliance rules against process data is a fundamental functionality for business process management. Over the years, the problem has been addressed for different types of process data, i.e., process models, process event data at runtime, and event logs representing historical execution. Several approaches have been proposed to tackle compliance checking over process logs. These approaches have been based on different data models and storage technologies including relational databases, graph databases, and proprietary formats. Graph-based encoding of event logs is a promising direction that turns several process analytics tasks into queries on the underlying graph. Compliance checking is one class of such analysis tasks.

In this paper, we argue that encoding log data as graphs alone is not enough to guarantee efficient processing of queries on this data. Efficiency is important due to the interactive nature of compliance checking. Thus, compliance checking would benefit from sub-linear scanning of the data. Moreover, as more data are added, e.g., new batches of logs arrive, the data size should grow sub-linearly to optimize both the space of storage and time for querying. We propose two encoding methods using graph representation, realized in Neo4J, and show the benefits of these encoding on a special class of queries, namely timed order compliance rules. Compared to a baseline encoding, our experiments show up to $5x$ speed up in the querying time as well as a $3x$ reduction in the graph size.

\end{abstract}

\section{Introduction}\label{sec:introduction}



Organizations strive to enhance their business processes to achieve several goals: increase customer satisfaction, gain more market share, reduce costs, and show adherence to regulations among other goals. Process mining techniques~\cite{processMiningBook2016} collectively help organizations achieve these goals by analyzing execution logs of organizations' information systems. Execution logs, a.k.a event logs, group events representing the execution of process steps into process instances (cases). Conformance checking~\cite{conformanceCheckingBook18}, in specific, provides techniques to analyze the deviation of the recorded behaviour against a behaviour represented as imperative process models or declarative rules~\cite{multiPerspectiveDECLARE16}.

Compliance checking~\cite{complianceRequirements2019} is a specialization of conformance checking in which event logs are checked against compliance rules that might restrict process behaviour w.r.t control flow, data, resources, and timing. Compliance rules represent interpretations of obligations either internal, e.g., policies, or external, e.g., legislation. Moreover, such rules are of a local nature. That is, they are not concerned with the end-to-end conformance of the process instance. Rather, they might refer to the execution ordering of a subset of the activities and their timing constraints. For example, in a ticketing system, there might be a rule that the time taken by creating a ticket and the first contact with the client should not exceed three hours. Compliance checking is an interactive and repetitive task by nature due to changes in the obligations. Compliance rules usually follow common patterns~\cite{elgammal2016formalizing,complianceRequirements2019}. The objective of compliance checking is to identify process instances that violate the rules. 

As compliance checking is an interactive process, event logs should be stored following data models that allow efficient access. Moreover, user-friendly domain-specific language, e.g., declarative query languages allow non-technical users to access and analyze the data. Recently, the graph data model has been investigated in the community to store and query event logs~\cite{processAtlas18,graphBasedProcessMining2020,multiDimlEventGrahDB21}. In this paper, we adopt the graph data model to represent event logs. Namely, we use the labeled property graph model. We propose two methods to represent event logs as graphs that can efficiently check compliance by means of translating compliance rules into queries. We address a special type (pattern) of compliance rules: \emph{order} patterns. However, our graph representation can address the rest of the patterns. We leave this discussion out due to space limitations. Namely, we make the following contributions:
\begin{inparaitem}
        \item We propose two graph representations of event logs that help efficiently check for compliance with \emph{order} rules,
        \item We empirically evaluate our method against the baseline graph representation and relational data models on a set of four real-life event logs. Our experiments show the scalability of our second encoding method,
         \item We discuss the improvements in the stored graph sizes and the simplification of the queries to check compliance. Overall, the compliance checking, i.e., querying time is improved by $3x$ to $5x$ whereas the sizes of the graphs are \texttildelow $3x$ reduced compared to the baseline method.
\end{inparaitem}

The rest of this paper is organized as follows: Section~\ref{sec:background} briefly discusses some of the background concepts and techniques that are used throughout the paper. Section~\ref{sec:proposed:approach} presents our approach. In Section \ref{sec:eval}, we evaluate the proposed approach against the existing one. Related work is discussed in Section
\ref{sec:related:work}. 


\section{Background}\label{sec:background}

\subsection{Events, Traces, Logs, and Graphs}\label{sub:sec:background:definitions}

We formalize the concepts of events, traces, logs and graphs to help in understanding the formalization introduced later in the paper.

\begin{definition}[Event]\label{def:event} An event $e$ is a tuple $(a_1,a_2,\dots, a_n)$ where $a_i$ is an attribute value drawn from a respective domain $a_i \in D_i$. At least three domains and their respective values must be defined for each event $e$: $D_c$, the set of case identifiers, $D_a$, the set of activity identifiers, and $D_t$, the set of timestamps. We denote these properties as $e.c$, $e.a$, and $e.t$ respectively. Other properties and domains are optional such as $D_r$, the resources who perform the tasks, $D_l$, the lifecycle phase of the activity.\end{definition}

We reserve the first three properties in the event tuple to reflect the case, the activity label, and the timestamp properties.

\begin{definition}[Trace]\label{def:trace} A trace is a finite sequence of events $\sigma = \langle e_1, e_2,\dots, e_m\rangle$ where $e_i$ is an event, $1 \leq i \leq m$ is a unique position for the event that identifies the event $e_i$ in $\sigma$ and explicitly positions it, and for any $e_i, e_j \in \sigma: e_i.c = e_j.c$
\end{definition}

\begin{definition}[Event log]\label{def:event:log} An event log is a finite sequence of events $\mathcal{L} = \langle e_1, e_2,\dots, e_m\rangle$ where events are ordered by their timestamps for any $e_i$ and $e_{i+1}: e_i.t \leq e_{i+1}.t$.
\end{definition}

In general, graph data models can be classified into two major groups~\cite{knowledgeGraph2022}: directed edge-labeled graphs, e.g., RDF, and labeled property graphs. In the context of this paper, we are interested in labeled property graphs as they provide a richer model that represents the same data in a smaller graph size.

\begin{definition}[Labeled property graph]\label{def:lpg} Let $L$, $K$, and $V$ be the sets of labels, keys, and values, respectively. A labeled property graph $\mathcal{G} = (N,E, label, prop)$ tuple, where $N$ is a non-empty set of nodes, $E\subseteq N\times N$ is the set of edges. $label: (N \cup E) \rightarrow 2^L$ is a labeling function to nodes and edges. $prop: (N \cup E) \times K \rightarrow V$ is a function that assigns key-value pairs to either nodes or edges.\end{definition}

When mapping from logs to graphs, we assume overloadings of a function $node()$ that identifies the corresponding node in the graph to the input parameter of the function. For instance for an event $e$, Definition~\ref{def:event}, $node(e)$ returns the corresponding node $n$ in $\mathcal{G}.N$ that represents the encoding of $e$. Similarly, for the case identifier $e.c$, $node(e.c)$ returns the node that corresponds to the case in the graph. Finally, for the activity label $e.a$, $node(e.a)$ returns the node that corresponds to the respective activity label.


\subsection{Compliance Patterns}\label{sec:compliance:patterns}

\begin{figure}
    \centering
    \includegraphics[width=\linewidth]{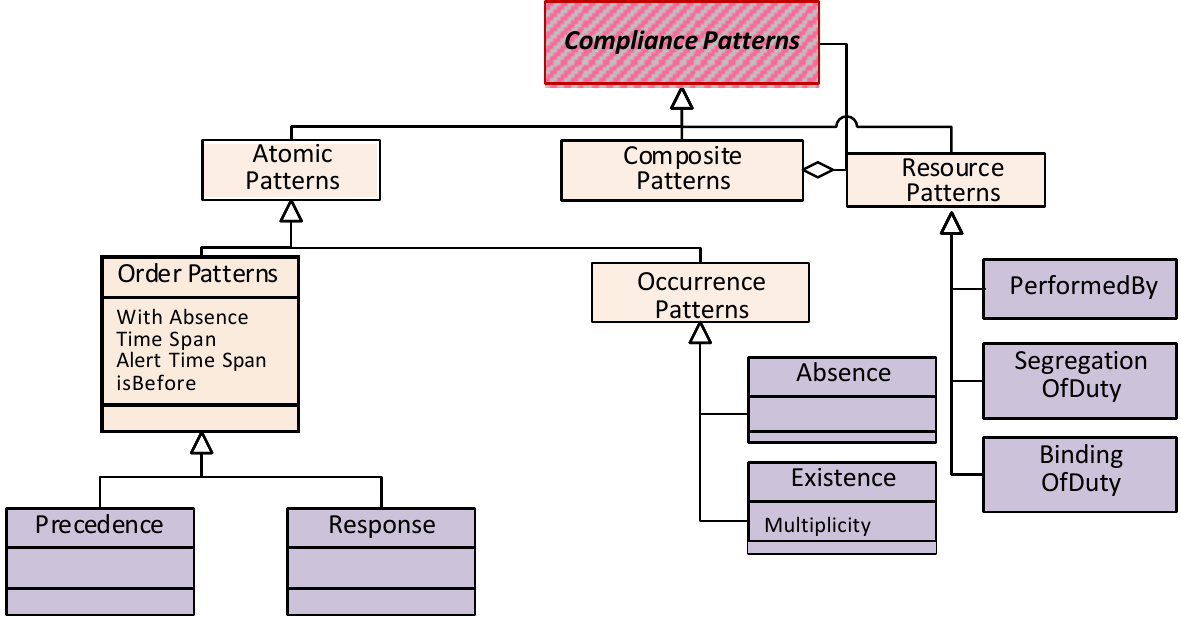}
    \caption{Categorization of compliance patterns}
    \label{fig:compliance:patterns}
 \end{figure}

A trace is compared to a process model in traditional conformance checking~\cite{conformanceCheckingBook18} to quantify the deviation between the required behaviour (the model) and the observed behaviour (the trace). Although multi-perspective techniques have been developed~\cite{conformanceCheckingSurvey19}, most conformance checking refers to control flow aspects of process execution. To compute deviations, these techniques need that the process models be enriched with resource, temporal, and data constraints. Moreover, the observed deviations are holistic on an end-to-end trace level. In many circumstances, checking deviations at a finer granularity, such as on the level of activities may be required, e.g., absence, existence, or pairs of activities, such as co-existence, mutual exclusion, and temporal and resource versions thereof. Such finer granularity checks are referred to as compliance checking, and \emph{compliance patterns} are used for categorizing the types of compliance requirements~\cite{elgammal2016formalizing,CMF15,complianceRequirements2019}. According to~\cite{elgammal2016formalizing}, a classification of patterns for business process compliance is shown in  Figure~\ref{fig:compliance:patterns}. 

Occurrence patterns are concerned with activities having been executed (\emph{Existence}) or not (\emph{Absence}) within a process instance. Order patterns are concerned with the execution order between pairs of activities. The \emph{Response} pattern (e.g., \emph{Response(A, B)}) states that if the execution of activity $A$ is observed at some point in a process instance, the execution of activity $B$ must be observed in some future point of the same case before the process instance is terminated. A temporal window can further restrict these patterns. For instance, we need to observe $B$ after $A$ in no more than a certain amount of time. Alternatively, we need to observe $B$ after observing $A$, where at least a certain amount of time have elapsed. Definition~\ref{def:response} formalizes the \emph{Response} pattern. 

\begin{definition}\textbf{Response --}\label{def:response}
Given two activities $A$ and $B$ and a trace $\trace = \langle \event_1, \event_2, \dots, \event_n \rangle$, $\trace \in \elog$, we say that $\trace \models Response(A,B, \Delta t, \theta)$ if and only if $\forall e_i \in \trace: {e_i.a}=A~\exists~e_j \in \trace: {e_j.b}=B \wedge {e_i.t} \leq {e_j.t} \wedge ({e_j.t} - {e_i.t})~ \theta~ \Delta t$, where $\Delta t$ represents the time window between $A$ and $B$, and $\theta \in \{<, =, >, \leq, \geq \}$ represents when (e.g., after, before, or exactly at) we expect the observation of $B$ after $A$ with respect to $\Delta t$.
\end{definition}

Conversely, the \emph{Precedes} pattern (e.g., \emph{Precedes(A,B)}) states that if the execution of activity $B$ is observed at some point in the trace, $A$ must have been observed before (Definition~\ref{def:precedes}). 

\begin{definition}\textbf{Precedes --}\label{def:precedes}
Given two activities $A$ and $B$ and  a trace $\trace = \langle \event_1, \event_2, \dots, \event_n \rangle$, $\trace \in \elog$, we say that $\trace \models Precedes(A,B, \Delta t, \theta)$ if and only if $\forall e_i \in \trace: {e_i.b}=B~\exists~e_j \in \trace: {e_j.a}=A \wedge {e_j.t} \leq {e_i.t} \wedge ({e_j.t} - {e_i.t})~ \theta~ \Delta t$.  where $\Delta t$ represents the time window between $A$ and $B$ and $\theta \in \{<, =, >, \leq, \geq \}$ represents when (e.g., after, before, or exactly at) we expect the observation of $A$ before $B$ with respect to $\Delta t$.
\end{definition}

Note that we get the unrestricted form of both patterns by setting $\Delta t$ to  a very large value and $\theta$ is set to $\leq$. That is, in $Response(A,B,\infty,\leq)$, $B$ has to \emph{eventually} be observed after $A$ with no further restrictions on the time window.

Both patterns can be further restricted by so-called \emph{exclude} constraint~\cite{AwadWW11}. That is, between the observations of of $A$ and $B$, it is prohibited to observe any of the activities listed in the exclude constraint.


 Resource patterns are concerned with the constraints on performers who act on pairs of activities, such as separation of duty or bind of duty. Such patterns can be combined to form composite patterns using different types of logical operators. In this paper, we focus on order patterns and their temporal variants. 

When checking for compliance, analysts are interested in identifying process instances, i.e., cases that contain a violation, rather than those that are compliant. Therefore, it is common in the literature about compliance checking to use the term ``anti-pattern''~\cite{DBLP:conf/sac/AwadBESAS15,complianceAntiPattern2015,businessProcessModelAntiPattern2019}. In the rest of this paper, we refer to anti-patterns rather than patterns when presenting our approach to detect compliance violations over graph-encoded event logs.

\section{Related Work}\label{sec:related:work}




There is vast literature about the business process compliance checking domain. For our purposes, we focus on compliance checking over event logs; we refer to this as auditing. For more details, the reader can check the survey in~\cite{surveyComplianceChecking18}.

Auditing can be categorized in basic terms based on the perspective of the process, including control-flow, data, resources, or time. We can also split these categories based on the formalism and technology that underpins them. Agrawal et al.~\cite{tamingComplianceDB06} presented one of the first works on compliance auditing, in which process execution data is imported into relational databases and compliance is verified by recognizing anomalous behaviour. This is done by comparing so-called workflow graphs of the rules and the process and looking for deviations. Control-flow-related topics are covered by the technique.

Validating process logs against control-flow and resource-aware compliance requirements have been proposed while applying model checking techniques ~\cite{processMiningVerificationLTL05}.
In~\cite{auditing2.010}, the authors proposed an adaptation of compliance checking techniques for auditing reasons. For control-flow and temporal rules, Ramezani et al.~\cite{complianceCheckingAlignment12,complianceCheckingTemporalRequirements13} suggest alignment-based detection of compliance violations.  In~\cite{complianceCheckingResourceData14}, another alignment-based approach for resource-related compliance violations is presented. 

%
De Murillas et al.~\cite{connectingDBPM19} present a metamodel and toolset for extracting process-related data from operational systems logs, such as relational databases, and populating their metamodel. A relational database is used to hold their metamodel. The authors show how different queries can be translated into SQL. Using relational databases provides support for a wide range of queries against process data. However, such queries are complex (using nesting, joins, and unions). Another alternative for addressing queries following slice, dice, drill-down, and roll-up operators of data cubes is OLAP-like analysis of process data. As a result, several approaches to storing event data in so-called process cubes~\cite{relationalDWPM15,multiDimensionalPM15} have been developed. In \cite{QueryLangWorkflow2022}, the authors extend traditional OLAP-like analysis to allow richer set of ad-hoc queries. Relational databases have also been used for declarative process mining~\cite{declarativeMiningSQL16}, which can be seen as an option for checking logs against compliance rules. 

Compliance violations, i.e. anti-patterns can be checked by \texttt{Match\_Recognize} (MR), the ANSI SQL operator. MR verifies patterns as regular expressions, where the tuples of a table are the symbols of the string to search for matches within. MR runs linearly through the number of tuples in the table. In our case, the tuples are the events in the log. In practice, the operational time can be enhanced by parallelizing the processing, e.g., partitioning the tuples by the case identifier. Still, this does not change the linearity of the match concerning the number of tuples in the table. A recent work speeds up MR by using indexes in relational databases~\cite{indexAcceleratedPatternMatching21} for strict contiguity patterns, i.e., patterns where events are in strict sequence. Order compliance patterns frequently refer to eventuality rather than strict order, limiting the use of indexes to accelerate the matching process.

Storing and querying event data into an integrated graph-based data structure has also been investigated.
Esser et al.~\cite{multiDimlEventGrahDB21} provide a rich data model for multi-dimensional event data using labeled property graphs realized on Neo4j as a graph database engine. To check for compliance, the authors use path queries. Such queries suffer from performance degradation when the distance between activities in the trace gets longer and when the whole graph size gets larger. 


\section{Graph-encoded Event Logs for Efficient Compliance Checking}\label{sec:proposed:approach}


Graph representation of event logs is a promising approach for event logs analysis~\cite{processAtlas18}, especially for compliance checking~\cite{multiDimlEventGrahDB21}. This is due to the richness of this graph representation model, mature database engines supporting it, e.g., Neo4J~\footnote{\scriptsize\url{https://neo4j.com/}}, and the declarative style of the query languages embraced by such engines, e.g., Cypher~\footnote{\scriptsize Cypher for Neo4J is like SQL for relational databases.}. In this sense, compliance checking can be mapped to queries against the encoded log to identify violations. 

We show how encoding of the event log has a significant effect on the efficiency of answering compliance queries. We start from a baseline approach (Section~\ref{sub:sec:encode:multi:base}) and propose two graph encoding methods, Sections~\ref{sub:sec:encode:multi:pos} and~\ref{sub:sec:encode:unique:event}, that leverage the finite nature of event logs to store the same event log in a smaller graph and answer compliance queries faster.

Table~\ref{table3:samplelogwithposition} shows an excerpt of a log that serves as the input to the different encoding methods. In the ``Optional details'' columns, the ``StartTime'' and ``CompleteTime'' columns are converted to Unix timestamp.

\begin{table}[htb!]
    \centering
    \caption{Sample event log with additional attributes}
    \vspace{0.3 cm}
    \begin{tabular}{c|c|o|o|o|x}
        \hline
    	\textbf{C.ID} & \textbf{Activity} & \textbf{Resource} & \textbf{StartTime} & \textbf{CompleteTime} & \textbf{Position} \\
    	\hline 
        1 & A & Jack & 1612172052 & 1612373652 & 1 \\
	    1 & B & John & 1612360812 & 1612458012 & 2 \\
	    2 & A & Mark & 1609491612 & 1609866012 & 1 \\
	    1 & E & Smith & 1612602012 & 1612778412 & 3 \\
	    3 & A & George & 1614589212 & 1614682812  & 1 \\
	    2 & C & Albert & 1609678812 & 1609866012 & 2 \\
	    1 & D & Mark & 1612954800 & 1613131200 & 4 \\
	    2 & E & Smith & 1611838812 & 1612026012 & 3 \\
	    3 & E & Albert & 1614934800  & 1615374000 & 2 \\
	    3 & C & Jack & 1615107612 & 1615374012  & 3 \\
	    2 & D & John & 1612256400  & 1612346400 & 4 \\
	    3 & E & Mark & 1615539600 & 1615719600 & 4 \\
	    3 & D & George & 1615546812 & 1615640412 & 5 \\
	   \tikzmark{a1} \vdots & \vdots\tikzmark{a2} & \tikzmark{b1}\vdots & \vdots & \vdots\tikzmark{b2} & \tikzmark{c1}\hspace{2mm}\vdots\hspace{2mm} \tikzmark{c2} \\
    	\hline 
	\end{tabular}
	\begin{tikzpicture}[overlay, remember picture]
        \draw [decorate,decoration={brace,amplitude=10pt,mirror,raise=4pt}] (a1.west) --node[below=14pt]{Minimum details} (a2.east);
        \draw [decorate,decoration={brace,amplitude=10pt,mirror,raise=4pt}] (b1.west) --node[below=14pt]{Optional details} (b2.east);
        \draw [decorate,decoration={brace,amplitude=10pt,mirror,raise=4pt}] (c1.west) --node[below=14pt]{Added detail} (c2.east);
    \end{tikzpicture}
    
	 \vspace{7mm}
      \label{table3:samplelogwithposition}
\end{table}

\subsection{Baseline: Multi-dimensional Graph Modeling (BM)}\label{sub:sec:encode:multi:base}

Esser at al.~\cite{multiDimlEventGrahDB21} proposed a multi-dimensional graph data model to represent event logs. It uses labeled property graphs, cf. Definition~\ref{def:lpg}, for the representation. \emph{Multi-dimensionality } is proposed as a flexible definition of a \emph{case} notion. However, for the scope of this paper, we will stick to the traditional definition of the case identifier. Yet, this simplification does not limit our contribution. Our proposed encoding methods can be generalized to any case notion embraced in the context of a specific compliance checking practice.

\begin{figure}[hbt!]
	\centering
	\begin{subfigure}[b]{\textwidth}
	    \centering
	    \includegraphics[scale=0.5]{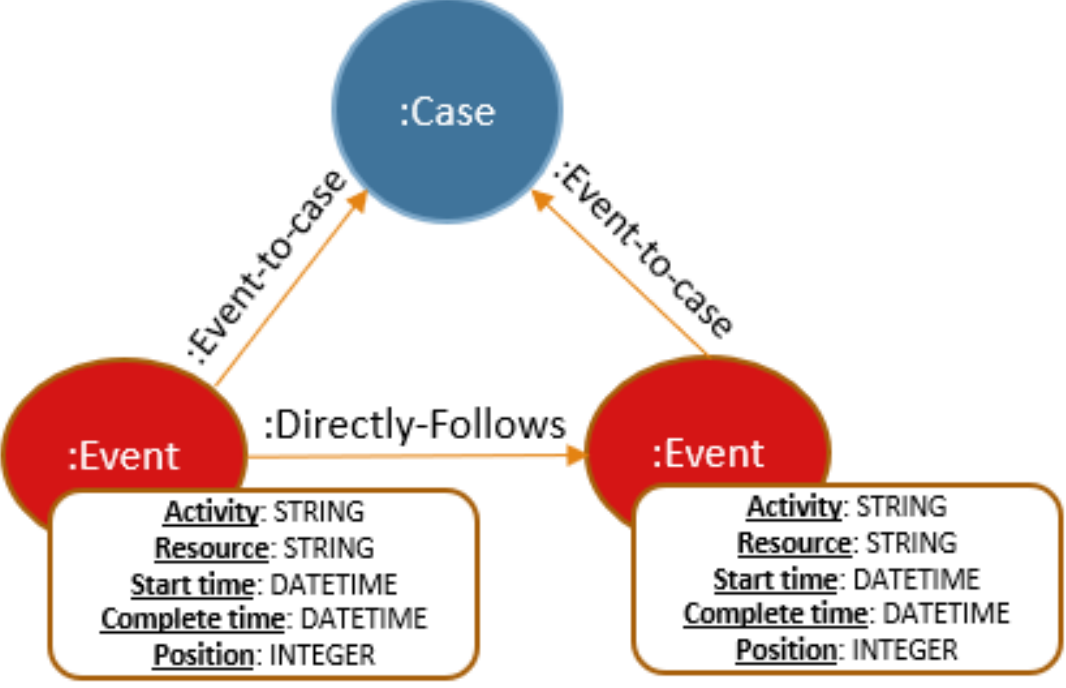}
	    \caption{Nodes, edges, labels, and properties}
	    \label{fig:encode:baseline}
	\end{subfigure}
	
	\begin{subfigure}[b]{\textwidth}
	    \centering
	    \includegraphics[scale=0.6]{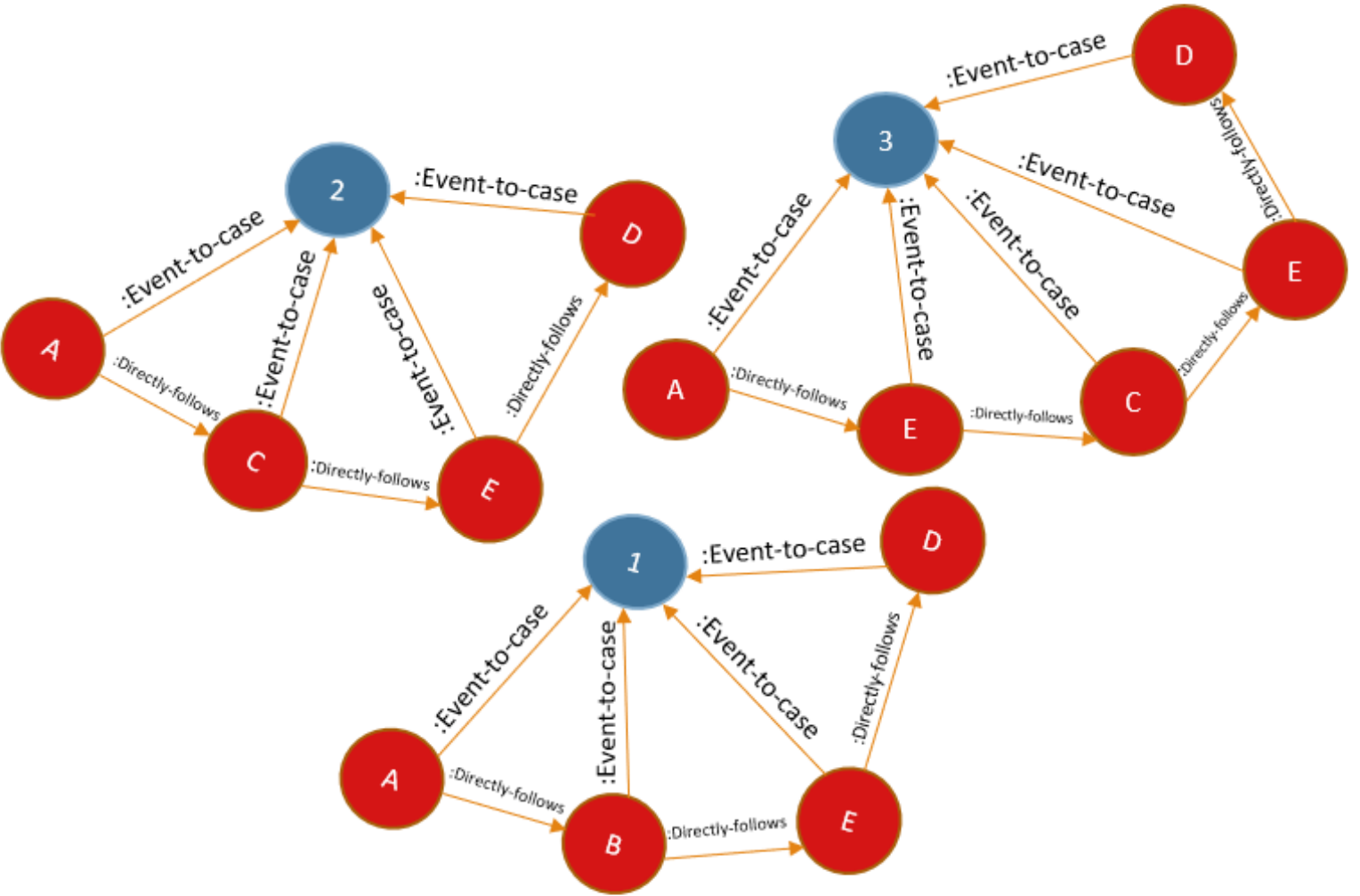}
	    \caption{Representation of the log excerpt in  Table~\ref{table3:samplelogwithposition}}
	    \label{fig:example:baseline:model}
	\end{subfigure}

\caption{Baseline graph representation} 
\label{fig:baseline:encodings}
\end{figure}

Events and cases constitute the nodes of the graph. Node types, i.e., events, cases, etc., are distinguished through labels. Edges represent either structural or behavioral relations. Structural relations represent event-to-case relations. Behavioral relations represent the execution order among events in the same case, referred to as directly-follows relationships. Activity labels, resource names, activity lifecycle status, and timestamps are modeled as properties of the event nodes. Similarly, case-level attributes are modeled as case node properties. Figure \ref{fig:encode:baseline} shows the representation of the Baseline graph.

Formally, for each log $\mathcal{L}$, cf. Definition~\ref{def:event:log}, a labeled property graph $\mathcal{G}$, cf. Definition~\ref{def:lpg}, is constructed by Esser et al.~\cite{multiDimlEventGrahDB21} approach as follows:

\begin{enumerate}
    \item Labels for the graph elements are constituted of four literals. Formally, $L=\{event, case, event\_to\_case, directly\_follows\}$,
    \item Keys for properties are the domain names from which values of the different event attributes are drawn. Formally, $K = \bigcup_{i=1}^{m}\{name(D_i)\} \cup \{ID\}$  ,
    \item For each unique case in the log, there is a node in the graph that is labeled as ``case'' and has a property ID that takes the value of the case identifier. Formally, $\forall c \in D_c~\exists~n\in \mathcal{G}.N: \{case\} \in label(n) \wedge prop (n, ID) = c $,
    \item For each event in the log, there is a node in the graph that is labeled as ``event'' and inherits this event's properties. Formally, $\forall e=(p_1,p_2,\dots,p_m)\\\in \mathcal{L}~\exists~n\in \mathcal{G}.N: \{event\} \in label(n) \wedge \forall_{2 \leq i \leq m} D_i: prop(n, name(D_i)) = p_i$,
    \item The structural relation between an event and its case is represented by a labeled relation. Formally, $\forall e\in \mathcal{L}, \exists~r\in \mathcal{G}.E: r=(node(e), node(e.c)) \wedge  \{event\_to\_case\} \in label(r)$,
    \item The behavioral relationship between a pair of successive events in a trace is represented by a labeled relation between their respective nodes. Formally, for $ e_i, e_{i+1}\in \tau~\exists~r\in \mathcal{G}.E: r=(node(e_i), node(e_{i+1})) \wedge  \{directly\_follows\} \in label(r)$
    
\end{enumerate}


In the following we adopt Cypher's notation to reflect on nodes, their labels, and their properties. We use the notation $:Label$ to refer the ``label'' of a graph element. For example, $:Event$ refers to the label ``event''. Figure~\ref{fig:example:baseline:model} visualizes the graph representation of the log excerpt given in Table~\ref{table3:samplelogwithposition} with the \textit{minimum details} columns.




Assume that we want to check a compliance rule that \emph{every execution of activity E must be preceded by an execution of activity B}, i.e., $Precedes(E,B,\infty,<)$. A violation of this rule is to find at least one execution of E that is not preceded by B from the beginning of a trace. The query in Listing~\ref{query1} expresses this anti-pattern using Cypher. Basically, the query first identifies the beginning of each trace \texttt{(start:Event\{activity:`A'\})}. Then the sequence of nodes constituting a path from each node of activity E to the start activity A, in the same case, is constructed. The path is constructed by traversing the transitive closure of the \emph{:Directly\_follows} relation, \texttt{path=(e1:Event\{activity:`E'\}<-[:Directly\_follows*]- (start))}. In case the path does not include any node whose activity property refers to B, as in line 3, a violation exists and this case is reported.
\vspace{-1 mm}
\setcounter{lstlisting}{0}
\renewcommand\thelstlisting{\arabic{lstlisting}}

\begin{lstlisting}[
language=SQL,
deletekeywords={IDENTITY},
deletekeywords={[2]INT},
morekeywords={clustered},
label={query1},
caption={Precedes anti-pattern query using the baseline encoding}]
Match (c1:Case) <-[:Event_to_case]- (start:Event{activity:'A'})
Match path = (e1:Event{activity:'E'}<-[:Directly_follows*]-(start)) 
where none (n in nodes(path) where n.activity='B')
return c.ID
\end{lstlisting}

Although the query is expressive and captures the semantics of the violation, it is expensive to evaluate. To answer the query, the processing engine has to scan the \emph{:Directly\_follows} linearly to resolve those nodes. Indeed, the class of order patterns, cf. Figure~\ref{fig:compliance:patterns}, will be expensive to evaluate under this graph encoding method. Another problem with this encoding is the linear growth of the graph size w.r.t the log size. In Figure~\ref{fig:example:baseline:model}, we can observe that each time an activity occurs in the log, a distinct node is created in the graph. 

In the following subsections, we propose a more succinct representation of the event log that improves both the space and time required to store the log and query it.

\subsection{Explicit Position Encoding  (EP)}\label{sub:sec:encode:multi:pos}

Many of the compliance patterns are concerned with the occurrence of activities in process execution and their ordering, cf. Figure~\ref{fig:compliance:patterns}. When checking such rules against event traces, we can exploit the finiteness of these traces and the positions of events within traces to simplify the queries and speed up their evaluation by utilizing indexes and skipping the linear scan of the \emph{:Directly\_follows} relation among events. So, we extend the baseline mapping by explicitly assigning a \emph{position} property to each event node. Table~\ref{table3:samplelogwithposition} has a highlighted column, tagged as \textit{added detail} column, where we assign each event to a position in the case (trace). For instance, the second row in Table~\ref{table3:samplelogwithposition} records that activity `B' has been the second activity to be executed in case $1$. Thus, the position property value is $2$. Likewise, the second row from the bottom of the table records that activity `E' has been executed as the fourth activity in case $3$. Therefore, the position property is $4$.




Having the position as a property of event nodes, we can simplify the query for the precedence anti-pattern to look as shown in Listing~\ref{query2}. We can observe that the check for ordering explicitly refers to the \emph{position} property of the event nodes without the expensive transitive closure traversal. The query looks for the occurrence of events of activity `E' where either there is no occurrence of activity `B' in the same case, or there is an occurrence of `B' which violates the order. This query can utilize indexes built on event properties, i.e., activity and position properties, to achieve sub-linear access to the nodes. Moreover, the query avoids the identification of the \emph{start} event of the trace, which was necessary for Listing~\ref{query1} to compute the transitive closure from \emph{start} to \emph{E}.

\begin{lstlisting}[
language=SQL,
deletekeywords={IDENTITY},
deletekeywords={[2]INT},
morekeywords={clustered},
label={query2},
caption={Precedes anti-pattern query utilizing event position property}]
Match (c2:Case),(b2:Event{activity:'B'}), (e2:Event{activity:'E'})
where (c2)<-[:Event_to_case]-(e2) and (not exists((c2)<- [:Event_to_case]-(b2)) or (exists((c2)<- [:Event_to_case]-(b2)) where e2.position < b2.position )))
return c.ID
\end{lstlisting}

Our encoding logic follows the same formalism shown in Section~\ref{sub:sec:encode:multi:base}, except for encoding the $directly\_follows$ relation as we add the explicit \emph{position} property to event nodes. The dropping of such a relation has a positive effect on the graph size.

\subsection{Unique Activities (UA)}\label{sub:sec:encode:unique:event}


Although the EP encoding method simplifies the processing of compliance queries, it inherits the linear growth of the graph size w.r.t the log size. To further limit the growth of the graph size, we modify the construction of the labeled property graph. This section's proposed encoding ensures a linear growth with the size of the set of activity labels, i.e., $D_a$. We generate a separate edge connecting a case node to the corresponding node representing the activity $a\in D_a$ and we add properties to the edges that reflect the position, timestamp, resource, etc. These properties of the events represent the execution of the activity in the respective case.

Formally, for each log $\mathcal{L}$, a labeled property graph $\mathcal{G}$ is constructed as follows:

\begin{enumerate}
    \item Labels for the graph elements are constituted of case, and activity labels. Formally, $\mathcal{L}=\{case, event\_to\_case\} \cup D_a$,
    \item Keys for properties are the domain names from which values of the different event attributes are drawn. Formally, $K = \bigcup_{i=1}^{m}\{name(D_i)\} \cup \{ID\}$ ,
    \item For each unique case in the log, there is a node in the graph that is labeled as ``case'' and has a property ID that takes the value of the case identifier. Formally, $\forall c \in D_c~\exists~n\in \mathcal{G}.N: \{case\} \in label(n) \wedge prop (n, ID) = c $,
    \item For each unique activity in the log, there is a node in the graph that is labeled as ``activity''. Formally,  $\forall a \in D_a~\exists~n\in \mathcal{G}.N: \{activity\} \in label(n)$,
    \item The structural relation between an event and its case is represented by a labeled relation between the activity node of the event's activity and the case node. Additionally, all event-level properties are mapped to properties on the edge. Formally, $\forall e=(p_1,p_2,\dots,p_m) \in \mathcal{L}~\exists~r\in \mathcal{G}.E: r=(node(e.a), node(e.c)) \wedge  \{event\_to\_case\} \in label(r) \wedge\\ \forall_{3 \leq i \leq m} D_i: prop(r, name(D_i)) = p_i$.

  
\end{enumerate}

\begin{figure}[hbt]
\centering
\includegraphics[scale=0.55]{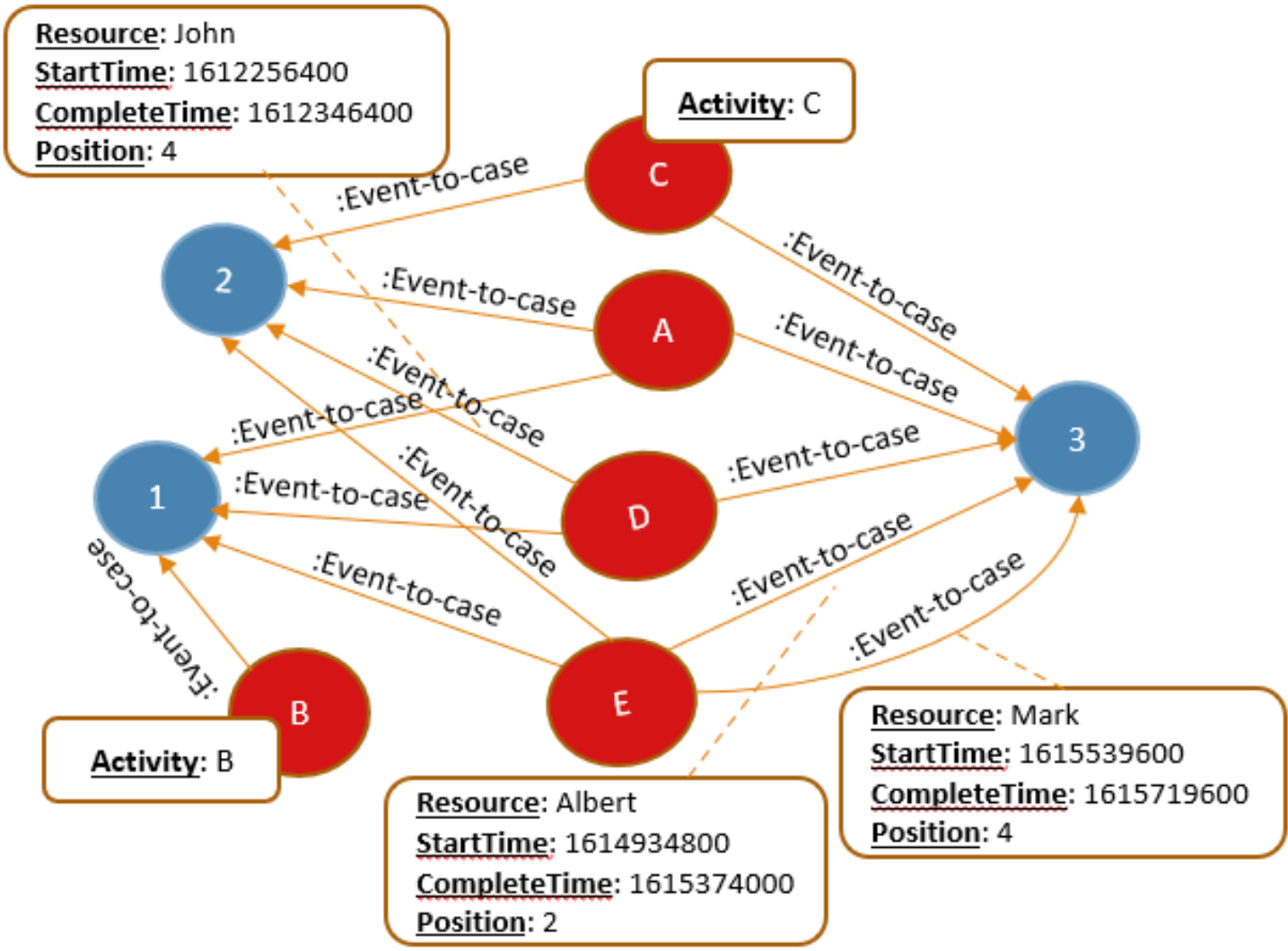}
\caption{Encoding of the log in Table~\ref{table3:samplelogwithposition} using UA method}
\label{fig:unique:activities:encodings}
\end{figure}

Figure~\ref{fig:unique:activities:encodings} visualizes the graph resulting from encoding the log excerpt in Table~\ref{table3:samplelogwithposition} using the unique activities method. For example, for activity \emph{E}, there is only one node and four different edges connecting to cases $1,~2,~and~3$. Two of these edges connect case $3$, as activity \emph{E} was executed twice in this case. 

\vspace{10mm}

Listing~\ref{query4} shows the modification on the \emph{Precedes} anti-pattern query. The query checks the ordering of the events using the \emph{position} property, which is accessed in Line 2. With this encoding, the graph size grows sub-linearly w.r.t the log. In fact, the size, i.e., the number of nodes in the graph, grows linearly w.r.t $|D_c|$, the set of case identifiers, which is significantly smaller than the number of events recorded in the log. However, for compliance checking purposes, queries mostly refer to activity labels to resolve nodes. 


\begin{lstlisting}[
language=SQL,
deletekeywords={IDENTITY},
deletekeywords={[2]INT},
morekeywords={clustered},
label={query4},
caption={Precedes anti-pattern query using unique activities encoding}]
Match (c3:Case)
where ((c3:Case)<-[r1:Event_to_Case]-(e3:Event{event:'E'}) and (c3:case) <-[r2:Event_to_Case]-(b3:Event{event:'B'}) and  r2.position > r1.position) or (not exists((c3)<-[:Event_to_Case]-(:Event{event:'B'}))
return c.ID
\end{lstlisting} 

Looking at the query in Listing~\ref{query4}, the database engine will handle this query by first binding variables $e3$ and $b3$ referring to activity labels `E' and `B', respectively. This binding would result in a single binding to explore for each variable. Next, relation variables $r1$ and $r2$ will be bound. The database engine can use the bindings of $e3$ and $b3$ and indexes on relation (edge) properties to prune the list of edges candidates for binding. Compared to the query in Listing~\ref{query1}, the node variable $e1$ will have as many bindings as there are events executed for activity `E'. The filter on activity label `B' cannot be used to prune the nodes traversed and stored in the $path$ variable. 

Comparing the query in Listing~\ref{query4} to the query in Listing~\ref{query2}, we observe that the former contains more variables to resolve, node and edge variables, whereas the latter contains node variables only. However, the candidate nodes to bind to variable $e3$ in the latter query will be bound to \emph{all} nodes for activity `E', compared to only a single binding in the former query. Next, variable $c2$ in Listing~\ref{query2}, will be bound to all case nodes linked to the nodes bound to $e2$. The binding of variable $b2$ in Listing~\ref{query2} will benefit from filtering with the activity label `B' to prune nodes. Further pruning can be achieved by the relation to the bound nodes to variables $c2$, the case node, and by using the position property of the variable $e2$. If `B' and `E' co-exist often in the same cases, the pruning via variable $c2$ is not helpful. Back to the query in Listing~\ref{query4}, the number of variables is bigger, but the candidates for node bindings are at most one, some activities may not have been executed at all. In turn, node bindings help reduce edge bindings, as we discussed earlier. Overall, the slightly larger number of variables does not affect the query processing time. This will be empirically proven when we evaluate these methods in the next section.

Theoretically, the UA method outperforms the EP method; the evaluation results empirically prove that. However, the EP method can still be applicable to log data that might have been already encoded with methods that encode the \emph{directly follows} relation, e.g., the BM method. In such situation, if migrating the data to the UA encoding is prohibitive, event nodes in the graph can be updated with the position property so that all queries related to comparing positions of the nodes within the trace, e.g., compliance queries, can be processed efficiently.


\section{Evaluation}\label{sec:eval}


This section reports the evaluation of the methods we proposed to encode event logs which are labeled property graphs. We compare our methods, EP and UA, against the baseline method BM. In addition, we compare to the storage of event logs in a relational table. The table consists of three columns to store the case ID, the activity, and the timestamp of the event. To detect compliance violations, we evaluate two approaches. The first uses common SQL operators such as joins and nested queries (NQ). The second uses the advanced \texttt{Match\_Recognize} (MR) operator. The event logs used for evaluation are described in Section~\ref{sub:sec:eval:dataset}. Implementation details and experimental setup are discussed in Section~\ref{sub:sec:eval:experiment:setting}. Finally, experimental results are discussed in Section~\ref{sub:sec:eval:discuss}.


\subsection{Data Sets}\label{sub:sec:eval:dataset}

We selected four real life logs: three logs from the BPI challenges to evaluate our experiments, namely: BPIC'12~\cite{vanDongen_2012}, BPIC'14\cite{vanDongen_2014}, BPIC'19~\cite{Fahland2021b} and the log namely: RTFMP~\cite{Reissner2022}. We considered these logs as they expose different characteristics as summarized in Table~\ref{tbl:char:real:logs}.  


\begin{table}
\vspace{-3mm}
\centering
\caption{Logs characteristics}
\vspace{0.2cm}
\begin{tabular}{|
>{\columncolor[HTML]{D6DCE4}} c|c|c|c|}
\hline
\textbf{Logs}  & {\cellcolor[HTML]{D6DCE4}\textbf{\#Traces}} & {\cellcolor[HTML]{D6DCE4}\textbf{\#Events}} & {\cellcolor[HTML]{D6DCE4}\textbf{\#Unique act.}} \\ \hline
\textbf{BPIC'12} & 13087           & 262200           & 24                         \\ \hline
\textbf{BPIC'14} & 41353           & 369485           & 9                           \\ \hline
\textbf{BPIC'19} & 220810          & 979942           & 8 \\ \hline

\textbf{RTFMP}  & 150370          & 561470          & 11\\ \hline
\end{tabular}
\label{tbl:char:real:logs}
\end{table}


\subsection{Implementation and Experimental Setup}\label{sub:sec:eval:experiment:setting}

We have implemented the encoding methods presented in sections~\ref{sub:sec:encode:multi:pos} and~\ref{sub:sec:encode:unique:event} using Neo4j version 4.3.1 as a graph database and Cypher to query the logs. Neo4J was instantiated with the default configuration of $1~GB$ of heap maximum size and a page cache size of $512~MB$. For the relational database, we used a docker image of Oracle 12c as it implements the \texttt{Match\_Recognize} operator. The instance was given main memory of $1~GB$ of RAM with default configurations of the storage engine. The experiments were run on a laptop running Windows 10 64-bit with an Intel Core i7 processor and $16~GB$ of RAM. 

To check for compliance, we prepared \textit{order patterns} queries, i.e. \emph{Response}, \emph{Precedes} and \emph{Exclude} anti-patterns. For each pattern, we created two variants. The two variants enforce a time limit with an upper and lower bound on the time limit, i.e., $Precedes(A,B, \Delta t, <)$, $Response(A,B, \Delta t, <)$ and $Exclude(A,B,C, \Delta t, <)$, respectively. For instance, in the case of a response query, we have $Response(A,B, \Delta t, <)$ and $Response(A,B, \Delta t, >)$. Therefore, in total, we have six queries for each log. The actual values for $A$, $B$ , $C$ and $\Delta t$ vary depending on the log. The anti-pattern queries for each variant are translated to Cypher and also to SQL for the respective encoding method to test. All the details for the rules (queries) variants and run details of experiments are available on Github~\footnote{\scriptsize\url{https://github.com/nesmayoussef/Graph-Encoded-Methods}}.

\subsection{Results and Discussion}\label{sub:sec:eval:discuss}

In the first experiment, we report on the loading time of the logs following the respective encoding, i.e. loading into Neo4J and the relational database (RDB). For each log, we report the loading time for the encoding methods and also, the number of nodes and edges created in the graph database, Table~\ref{tab:dom:knowledge:logs}. To unify the loading steps in graph database, we make two passes on the log. The first pass creates the nodes and the second pass creates the edges between these nodes. The rationale behind this separation is that Neo4J is a transactional database and it uses locks whenever nodes are concurrently accessed. As we execute the commands to add nodes and edges, Neo4J will use the available threads to execute the commands concurrently. To factor out the effects of locking on loading time, we load the nodes and edges in separate steps. 
\begin{table}[htbp]
\centering
\caption{Loading time (in seconds) for each encoding method. [LT: Loading Time, \# N: number of nodes, \# E: number of edges]}
\vspace{0.3 cm}
\label{tab:dom:knowledge:logs}
\begin{tabular}{|
>{\columncolor[HTML]{D6DCE4}}l l|ccc|ccc|ccc|c|}
\hline
\multicolumn{2}{|c|}{\cellcolor[HTML]{D6DCE4}\textbf{Methods}} &
  \multicolumn{3}{c|}{\cellcolor[HTML]{D6DCE4}\textbf{BM}} &
  \multicolumn{3}{c|}{\cellcolor[HTML]{D6DCE4}\textbf{EP}} &
  \multicolumn{3}{c|}{\cellcolor[HTML]{D6DCE4}\textbf{UA}} & \multicolumn{1}{c|}{\cellcolor[HTML]{D6DCE4}\textbf{RDB}}\\ \hline
\multicolumn{1}{|l|}{\cellcolor[HTML]{D6DCE4}\textbf{Logs}} &
  \textbf{\# Cases} &
  \multicolumn{1}{c|}{\textbf{LT}} &
  \multicolumn{1}{c|}{\textbf{\# N}} &
  \textbf{\# E} &
  \multicolumn{1}{c|}{\textbf{LT}} &
  \multicolumn{1}{c|}{\textbf{\# N}} &
  \textbf{\# E} &
  \multicolumn{1}{c|}{\textbf{LT}} &
  \multicolumn{1}{c|}{\textbf{\# N}} &
  \textbf{\# E} & \multicolumn{1}{c|}{\textbf{LT}} \\ \hline
\multicolumn{1}{|l|}{\multirow{-1}{*}{\cellcolor[HTML]{D6DCE4}\textbf{BPIC'12}}} &
  \textbf{13087} &
  \multicolumn{1}{c|}{16} &
  \multicolumn{1}{c|}{177,597} &
  315,933 &
  \multicolumn{1}{c|}{12} &
  \multicolumn{1}{c|}{177,597} &
  164,510 &
  \multicolumn{1}{c|}{8.5} &
  \multicolumn{1}{c|}{13,111} &
  164,510 & \multicolumn{1}{c|}{1641}\\ \hline
\multicolumn{1}{|l|}{\cellcolor[HTML]{D6DCE4}} &
  \textbf{15000} &
  \multicolumn{1}{c|}{13} &
  \multicolumn{1}{c|}{148,883} &
  252,766 &
  \multicolumn{1}{c|}{10} &
  \multicolumn{1}{c|}{148,883} &
  133,883 &
  \multicolumn{1}{c|}{7} &
  \multicolumn{1}{c|}{15,009} &
  133,883 & \multicolumn{1}{c|}{890}\\ 
\multicolumn{1}{|l|}{\multirow{-2}{*}{\cellcolor[HTML]{D6DCE4}\textbf{BPIC'14}}} &
  \textbf{41353} &
  \multicolumn{1}{c|}{\textbf{\textemdash}} &
  \multicolumn{1}{c|}{\textbf{\textemdash}} &\multicolumn{1}{c|}{\textbf{\textemdash}}
   &
  \multicolumn{1}{c|}{\textbf{\textemdash}} &
  \multicolumn{1}{c|}{\textbf{\textemdash}} & \multicolumn{1}{c|}{\textbf{\textemdash}}
   &
  \multicolumn{1}{c|}{12} &
  \multicolumn{1}{c|}{41,362} &
  369,480 & \multicolumn{1}{c|}{2447} \\ \hline
\multicolumn{1}{|l|}{\cellcolor[HTML]{D6DCE4}} &
  \textbf{25000} &
  \multicolumn{1}{c|}{13} &
  \multicolumn{1}{c|}{135,933} &
  196,866 &
  \multicolumn{1}{c|}{10} &
  \multicolumn{1}{c|}{135,933} &
  110,933 &
  \multicolumn{1}{c|}{6} &
  \multicolumn{1}{c|}{25,008} &
  110,933 & \multicolumn{1}{c|}{960}\\ 
\multicolumn{1}{|l|}{\multirow{-2}{*}{\cellcolor[HTML]{D6DCE4}\textbf{BPIC'19}}} &
  \textbf{220810} &
  \multicolumn{1}{c|}{\textbf{\textemdash}} &
  \multicolumn{1}{c|}{\textbf{\textemdash}} & \multicolumn{1}{c|}{\textbf{\textemdash}}
   &
  \multicolumn{1}{c|}{\textbf{\textemdash}} &
  \multicolumn{1}{c|}{\textbf{\textemdash}} & \multicolumn{1}{c|}{\textbf{\textemdash}}
   &
  \multicolumn{1}{c|}{60} &
  \multicolumn{1}{c|}{220,818} &
  976,994 & \multicolumn{1}{c|}{8153}\\ \hline
\multicolumn{1}{|l|}{\cellcolor[HTML]{D6DCE4}} &
  \textbf{50000} &
  \multicolumn{1}{c|}{21} &
  \multicolumn{1}{c|}{236,633} &
  323,266 &
  \multicolumn{1}{c|}{12} &
  \multicolumn{1}{c|}{236,633} &
  186,633 &
  \multicolumn{1}{c|}{10} &
  \multicolumn{1}{c|}{50,011} &
  186,633 & \multicolumn{1}{c|}{1250}\\ 
\multicolumn{1}{|l|}{\multirow{-2}{*}{\cellcolor[HTML]{D6DCE4}\textbf{RTFMP}}} &
  \textbf{150370} &
  \multicolumn{1}{c|}{\textbf{\textemdash}} &
  \multicolumn{1}{c|}{\textbf{\textemdash}} & \multicolumn{1}{c|}{\textbf{\textemdash}}
   &
  \multicolumn{1}{c|}{\textbf{\textemdash}} &
  \multicolumn{1}{c|}{\textbf{\textemdash}} &\multicolumn{1}{c|}{\textbf{\textemdash}}
   &
  \multicolumn{1}{c|}{33} &
  \multicolumn{1}{c|}{150,381} &
  561,440 & \multicolumn{1}{c|}{3731} \\ \hline
\end{tabular}
\end{table}

Table~\ref{tab:dom:knowledge:logs} reports the outcomes of the first experiment. In our experiments, for large logs, the BM and EP methods, Neo4J crashed with an out of memory error due to the large amount of data. This is the case for the BPIC'14, '19, and the RTFMP logs. We have examined several subsets of these logs and the number of cases reported in the table corresponds to the maximum size that could be loaded using the Neo4J configuration we mentioned earlier. For the UA and RDB encoding, all the data are loaded into the database for the full log sizes. For the common log sizes among the encoding methods,  graph-based encoding shows superiority to the relational database when loading the data. Additionally, within graph encoding methods, UA is the fastest due to the smaller number of nodes and edges compared to the other two graph encoding methods, yet capturing the same behaviour. In many cases, the loading time of the UA method is $2x$ faster than the baseline and $150x$ faster than RDB. 


Turning to graph sizes, we can observe the reduction of their sizes when encoding with the EP and UA methods. The former reduces the number of edges, as it does not store edges for the \emph{directly follows} relation. The latter reduces the size of the nodes while keeping the same number of edges as the former. However, the degree of the nodes, i.e., the number of incoming and outgoing edges, is different. Table \ref{tab:average:degree} reports the average degree of the nodes for the EP and UA methods. We observe that the average degree for the nodes in the UA is higher. The value of such a high degree, given how Neo4J stores its data~\footnote{\scriptsize\url{https://neo4j.com/developer/kb/understanding-data-on-disk/}}, improves the localization of data access leading to faster query execution.

\begin{table}[htbp!]
\centering
\caption{Average degree of the nodes for EP and UA methods}
\label{tab:average:degree}
\vspace{0.2 cm}
\begin{tabular}{|
>{\columncolor[HTML]{D6DCE4}}l |c|c|c|}
\hline
\multicolumn{1}{|c|}{\cellcolor[HTML]{D6DCE4}\textbf{Logs}} &
  \cellcolor[HTML]{D6DCE4}\textbf{\# Cases} &
  \cellcolor[HTML]{D6DCE4}\textbf{EP} &
  \cellcolor[HTML]{D6DCE4}\textbf{UA} \\ \hline
\textbf{BPIC'12} & \textbf{13087} & 0.93 & 12.54 \\ \hline
\textbf{BPIC'14} & \textbf{15000} & 0.89 & 8.92  \\ \hline
\textbf{BPIC'19} & \textbf{25000} & 0.82 & 4.5   \\ \hline
\textbf{RTFMP}   & \textbf{50000} & 0.78 & 3.73  \\ \hline
\end{tabular}
\end{table}

In the second experiment, we run the four compliance anti-pattern queries against the respective logs, two for response and two for precedes. We report the execution time of the queries for the different encoding methods in Table~\ref{tab:datasets:precedence} and Table~\ref{tab:datasets:response}.

\begin{table}[htbp!]
\centering
\caption{ Execution time (in milli sec.) for the variants of the \emph{Precedes} queries [W: Within time window, B: Before time window]}
\label{tab:datasets:precedence}
\vspace{0.2 cm}
\begin{tabular}{|cl|ll|ll|ll|ll|ll|}
\hline
\rowcolor[HTML]{D6DCE4} 
\multicolumn{2}{|c|}{\cellcolor[HTML]{D6DCE4}\textbf{Methods}} &
  \multicolumn{2}{c|}{\cellcolor[HTML]{D6DCE4}\textbf{NQ}} &
  \multicolumn{2}{c|}{\cellcolor[HTML]{D6DCE4}\textbf{MR}} &
  \multicolumn{2}{c|}{\cellcolor[HTML]{D6DCE4}\textbf{BM}} &
  \multicolumn{2}{c|}{\cellcolor[HTML]{D6DCE4}\textbf{EP}} &
  \multicolumn{2}{c|}{\cellcolor[HTML]{D6DCE4}\textbf{UA}} \\ \hline
\rowcolor[HTML]{D6DCE4} 
\multicolumn{1}{|c|}{\cellcolor[HTML]{D6DCE4}\textbf{Log}} &
  \multicolumn{1}{c|}{\cellcolor[HTML]{D6DCE4}\textbf{\# Cases}} &
  \multicolumn{1}{c|}{\cellcolor[HTML]{D6DCE4}\textbf{W}} &
  \multicolumn{1}{c|}{\cellcolor[HTML]{D6DCE4}\textbf{B}} &
  \multicolumn{1}{c|}{\cellcolor[HTML]{D6DCE4}\textbf{W}} &
  \multicolumn{1}{c|}{\cellcolor[HTML]{D6DCE4}\textbf{B}} &
  \multicolumn{1}{c|}{\cellcolor[HTML]{D6DCE4}\textbf{W}} &
  \multicolumn{1}{c|}{\cellcolor[HTML]{D6DCE4}\textbf{B}} &
  \multicolumn{1}{c|}{\cellcolor[HTML]{D6DCE4}\textbf{W}} &
  \multicolumn{1}{c|}{\cellcolor[HTML]{D6DCE4}\textbf{B}} &
  \multicolumn{1}{c|}{\cellcolor[HTML]{D6DCE4}\textbf{W}} &
  \multicolumn{1}{c|}{\cellcolor[HTML]{D6DCE4}\textbf{B}} \\ \hline
\multicolumn{1}{|l|}{\cellcolor[HTML]{D6DCE4}\textbf{BPIC'12}} &
  \textbf{13087} &
  \multicolumn{1}{l|}{177} &
  184 &
  \multicolumn{1}{l|}{723} &
  571 &
  \multicolumn{1}{l|}{138} &
  74 &
  \multicolumn{1}{l|}{81} &
  127 &
  \multicolumn{1}{l|}{59} &
  52 \\ \hline
\multicolumn{1}{|c|}{\cellcolor[HTML]{D6DCE4}} &
  \textbf{15000} &
  \multicolumn{1}{l|}{153} &
  522 &
  \multicolumn{1}{l|}{633} &
  432 &
  \multicolumn{1}{l|}{85} &
  253 &
  \multicolumn{1}{l|}{39} &
  89 &
  \multicolumn{1}{l|}{29} &
  127 \\ \cline{2-12} 
\multicolumn{1}{|c|}{\multirow{-2}{*}{\cellcolor[HTML]{D6DCE4}\textbf{BPIC'14}}} &
  \textbf{41373} &
  \multicolumn{1}{l|}{519} &
  367 &
  \multicolumn{1}{l|}{1759} &
  1202 &
  \multicolumn{1}{l|}{\textbf{\textemdash}} & \textbf{\textemdash}
   &
  \multicolumn{1}{l|}{\textbf{\textemdash}} & \textbf{\textemdash}
   &
  \multicolumn{1}{l|}{37} &
  24 \\ \hline
\multicolumn{1}{|c|}{\cellcolor[HTML]{D6DCE4}} &
  \textbf{25000} &
  \multicolumn{1}{l|}{128} &
  277 &
  \multicolumn{1}{l|}{634} &
  1006 &
  \multicolumn{1}{l|}{68} &
  81 &
  \multicolumn{1}{l|}{76} &
  52 &
  \multicolumn{1}{l|}{52} &
  39 \\ \cline{2-12} 
\multicolumn{1}{|c|}{\multirow{-2}{*}{\cellcolor[HTML]{D6DCE4}\textbf{BPIC'19}}} &
  \textbf{220810} &
  \multicolumn{1}{l|}{910} &
  1537 &
  \multicolumn{1}{l|}{5770} &
  9154 &
  \multicolumn{1}{l|}{\textbf{\textemdash}} & \textbf{\textemdash}
   &
  \multicolumn{1}{l|}{\textbf{\textemdash}} & \textbf{\textemdash}
   &
  \multicolumn{1}{l|}{101} &
  79 \\ \hline
\multicolumn{1}{|l|}{\cellcolor[HTML]{D6DCE4}} &
  \textbf{50000} &
  \multicolumn{1}{l|}{185} &
  537 &
  \multicolumn{1}{l|}{477} &
  1106 &
  \multicolumn{1}{l|}{83} &
  137 &
  \multicolumn{1}{l|}{81} &
  95 &
  \multicolumn{1}{l|}{49} &
  61 \\ \cline{2-12} 
\multicolumn{1}{|l|}{\multirow{-2}{*}{\cellcolor[HTML]{D6DCE4}\textbf{RTFMP}}} &
  \textbf{150370} &
  \multicolumn{1}{l|}{406} &
  1147 &
  \multicolumn{1}{l|}{1447} &
  3352 &
  \multicolumn{1}{l|}{\textbf{\textemdash}} & \textbf{\textemdash}
   &
  \multicolumn{1}{l|}{\textbf{\textemdash}} &\textbf{\textemdash}
  &
  \multicolumn{1}{l|}{193} &
  127 \\ \hline
\end{tabular}
\end{table}

\begin{table}[htbp!]
\centering
\caption{Execution time (in milli sec.) for the variants of the \emph{Response} queries [W: Within time window,  A: After time window]}
\label{tab:datasets:response}
\vspace{0.2 cm}
\begin{tabular}{|cc|cc|cc|cc|cc|cc|}
\hline
\rowcolor[HTML]{D6DCE4} 
\multicolumn{2}{|c|}{\cellcolor[HTML]{D6DCE4}\textbf{Methods}} &
  \multicolumn{2}{c|}{\cellcolor[HTML]{D6DCE4}\textbf{NQ}} &
  \multicolumn{2}{c|}{\cellcolor[HTML]{D6DCE4}\textbf{MR}} &
  \multicolumn{2}{c|}{\cellcolor[HTML]{D6DCE4}\textbf{BM}} &
  \multicolumn{2}{c|}{\cellcolor[HTML]{D6DCE4}\textbf{EP}} &
  \multicolumn{2}{c|}{\cellcolor[HTML]{D6DCE4}\textbf{UA}} \\ \hline
\rowcolor[HTML]{D6DCE4} 
\multicolumn{1}{|c|}{\cellcolor[HTML]{D6DCE4}\textbf{Log}} &
  \textbf{\# Cases} &
  \multicolumn{1}{c|}{\cellcolor[HTML]{D6DCE4}\textbf{W}} &
  \textbf{A} &
  \multicolumn{1}{c|}{\cellcolor[HTML]{D6DCE4}\textbf{W}} &
  \textbf{A} &
  \multicolumn{1}{c|}{\cellcolor[HTML]{D6DCE4}\textbf{W}} &
  \textbf{A} &
  \multicolumn{1}{c|}{\cellcolor[HTML]{D6DCE4}\textbf{W}} &
  \textbf{A} &
  \multicolumn{1}{c|}{\cellcolor[HTML]{D6DCE4}\textbf{W}} &
  \textbf{A} \\ \hline
\multicolumn{1}{|c|}{\cellcolor[HTML]{D6DCE4}\textbf{BPIC'12}} &
  \textbf{13087} &
  \multicolumn{1}{c|}{171} &
  97 &
  \multicolumn{1}{c|}{953} &
  747 &
  \multicolumn{1}{c|}{12} &
  112 &
  \multicolumn{1}{c|}{11} &
  91 &
  \multicolumn{1}{c|}{4} &
  33 \\ \hline
\multicolumn{1}{|c|}{\cellcolor[HTML]{D6DCE4}} &
  \textbf{15000} &
  \multicolumn{1}{c|}{146} &
  121 &
  \multicolumn{1}{c|}{712} &
  425 &
  \multicolumn{1}{c|}{46} &
  71 &
  \multicolumn{1}{c|}{37} &
  61 &
  \multicolumn{1}{c|}{31} &
  38 \\ \cline{2-12} 
\multicolumn{1}{|c|}{\multirow{-2}{*}{\cellcolor[HTML]{D6DCE4}\textbf{BPIC'14}}} &
  \textbf{41373} &
  \multicolumn{1}{c|}{522} &
  281 &
  \multicolumn{1}{c|}{1979} &
  1181 &
  \multicolumn{1}{c|}{\textbf{\textemdash}} & \textbf{\textemdash}
   &
  \multicolumn{1}{c|}{\textbf{\textemdash}} & \textbf{\textemdash}
   &
  \multicolumn{1}{c|}{49} &
  56 \\ \hline
\multicolumn{1}{|c|}{\cellcolor[HTML]{D6DCE4}} &
  \textbf{25000} &
  \multicolumn{1}{c|}{152} &
  191 &
  \multicolumn{1}{c|}{931} &
  1267 &
  \multicolumn{1}{c|}{21} &
  29 &
  \multicolumn{1}{c|}{21} &
  21 &
  \multicolumn{1}{c|}{12} &
  15 \\ \cline{2-12} 
\multicolumn{1}{|c|}{\multirow{-2}{*}{\cellcolor[HTML]{D6DCE4}\textbf{BPIC'19}}} &
  \textbf{220810} &
  \multicolumn{1}{c|}{1029} &
  1521 &
  \multicolumn{1}{c|}{8461} &
  11523 &
  \multicolumn{1}{c|}{\textbf{\textemdash}} & \textbf{\textemdash}
   &
  \multicolumn{1}{c|}{\textbf{\textemdash}} & \textbf{\textemdash}
   &
  \multicolumn{1}{c|}{82} &
  47 \\ \hline
\multicolumn{1}{|c|}{\cellcolor[HTML]{D6DCE4}} &
  \textbf{50000} &
  \multicolumn{1}{c|}{223} &
  402 &
  \multicolumn{1}{c|}{932} &
  1433 &
  \multicolumn{1}{c|}{37} &
  59 &
  \multicolumn{1}{c|}{33} &
  32 &
  \multicolumn{1}{c|}{24} &
  17 \\ \cline{2-12} 
\multicolumn{1}{|c|}{\multirow{-2}{*}{\cellcolor[HTML]{D6DCE4}\textbf{RTFMP}}} &
  \textbf{150370} &
  \multicolumn{1}{c|}{585} &
  1158 &
  \multicolumn{1}{c|}{2824} &
  4345 &
  \multicolumn{1}{c|}{\textbf{\textemdash}} & \textbf{\textemdash}
   &
  \multicolumn{1}{c|}{\textbf{\textemdash}} & \textbf{\textemdash}
   &
  \multicolumn{1}{c|}{26} &
  52 \\ \hline
\end{tabular}
\end{table}

\begin{figure}[ht!]
	\centering
	\begin{subfigure}[b]{\textwidth}
	    \centering
	    \includegraphics[width=0.75\textwidth]{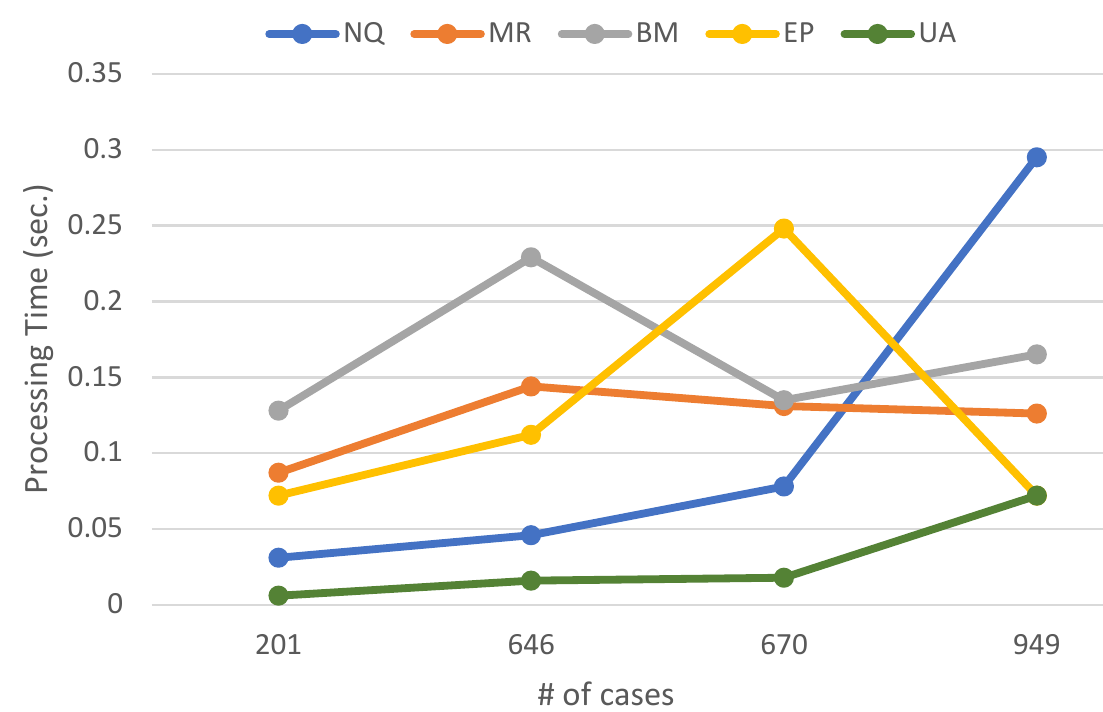}
	    \caption{No time window}
	    \label{fig:exclude:noTime}
	\end{subfigure}
	\begin{subfigure}[b]{\textwidth}
	    \centering
	    \includegraphics[width=0.75\textwidth]{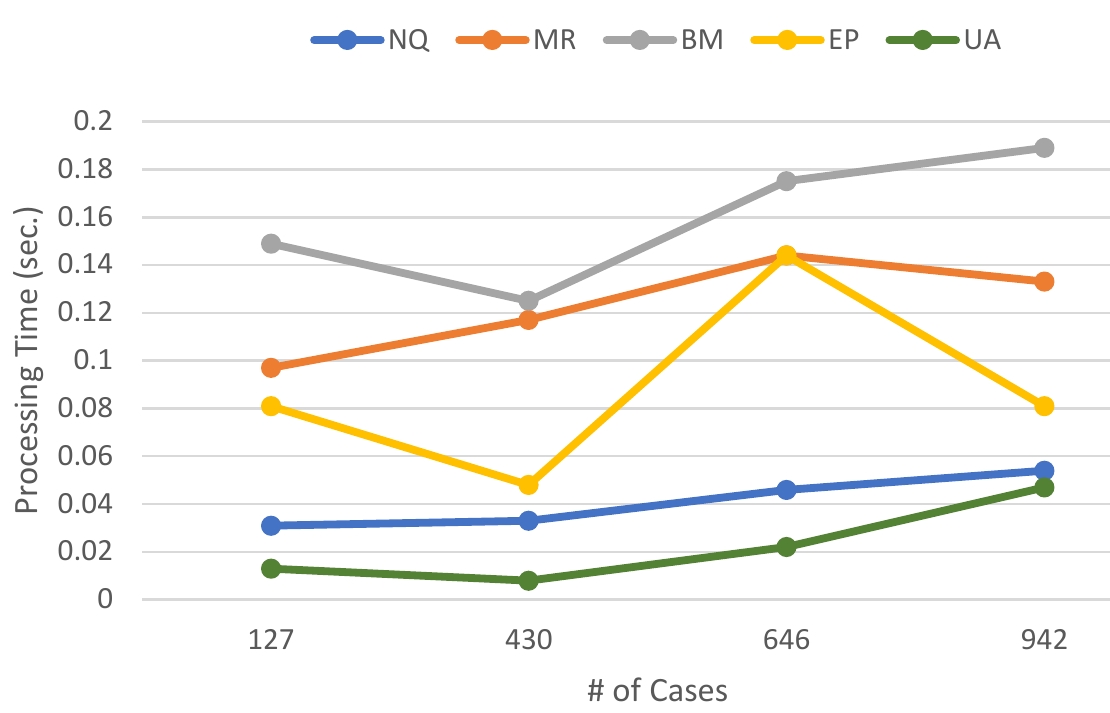}
	    \caption{With time window}
	    \label{fig:exclude:afterTime}
	\end{subfigure}
	
\caption{Comparing the results of Exclude queries} 
\label{fig:exclude:results}
\end{figure}


Tables~\ref{tab:datasets:precedence} and~\ref{tab:datasets:response} report the average execution time of the \emph{Precedes} and \emph{Response} anti-pattern queries, respectively. Overall, the execution time is reduced using the proposed encoding methods compared to the baseline method BM, NQ and MR. The magnitude of gain differs, though. 

For the precedence anti-patterns, in the case of the UA method, the reduction of execution time goes up to $10x$, as in the case of the BPIC'12 log for the Before time limit, B, query in Table~\ref{tab:datasets:precedence} compared to execution time of MR and NQ. In NQ, we use nested queries and self joins which leads the query engine to perform additional tasks to retrieve data. Using MR, the database has to scan all the records and match them to the non-deterministic finite automata (NFA) to check for matches. Comparing UA to the BM graph encoding, we still get an improvement in query time. The gain goes up to $2x$ as in the case of BPIC'14 log for the W query.

For the response anti-patterns, in BPIC'12 log, for within time window,W, MR and NQ perform worse than the other methods, Table \ref{tab:datasets:response}. The improvement in query time goes up to $238x$ comparing UA to MR. The lowest improvement is about $2x$, comparing UA to other graph encoding methods. Note that this gain is on a small subset of the log. It is not clear how fast the processing would be if the full log was loaded using the BM and EP methods. This is left as a future work when testing with higher hardware specification.

Comparing the EP method to the other methods, we still get a gain in performance in most of the log/query combinations. However, the magnitude of the gain is less compared to the UA method. The best gain compared to MR is $19x$ in the case of the BPIC'19 log for Before time window and $60x$ in the case of the BPIC'19 log for After time window, Table~\ref{tab:datasets:precedence} and Table~\ref{tab:datasets:response}, respectively. However, in few cases, EP performs worse than BM as shown in Table~\ref{tab:datasets:precedence}. For instance, in the BPIC'12 log, for the before time window, B,query. Although the difference is quite small, it could be explained by the fact that the queried activities are already in direct contiguity in the execution trace. This further affirms the limitation of low node degree in the EP method which minimizes localization of data access. 
 
 In the third experiment, we run \emph{Exclude} anti-pattern queries against BPIC'15~\cite{Fahland2021a} log. This log contains 1199 cases with a total of 52217 events and 398 unique events. We chose this log due to its huge number of unique events. Here, we empirically validate that the proposed method still gives the best execution time. This experiment was run four times with different activities and time window for the five encoding methods.
 
 Figures \ref{fig:exclude:noTime} and \ref{fig:exclude:afterTime} report the execution time of the queries, with and without time window, respectively. We show on the x-axis the query results sorted by the matching number of cases. Obviously, the UA method shows the best scalability as the number of matching cases (process instances) is a function in both the input log size and the anti-pattern query.
 
 Overall, the graph-based encoding of event logs shows superiority over the traditional relational database encoding. This aligns with recent directions to employ graph databases for process analytics~\cite{multiDimlEventGrahDB21}. Additionally, the UA encoding method we propose provides improvement in both query time and storage space against the baseline BM graph encoding method.


\section{Conclusion and Future Work}\label{sec:conclusion:future}

In this paper, we propose two graph-based encoding methods for event logs to efficiently check their compliance with timed order rules. The first encoding method enhances an encoding already proposed in the literature, whereas the second is new. Both methods enhance the order compliance checking time as well as reduce the size of the stored graphs. Our experimental evaluation empirically confirms the gain in both directions. In addition, our experimental evaluation compares with encoding logs in a relational table and employing traditional SQL as well as pattern matching operators MR to realize querying (checking) for violations.

A limitation of this work is that it has been evaluated using Neo4J only as a graph database engine. Unlike the relational model, the graph data models' representation and querying are not standardized and lack an algebraic basis. Thus, the gains in performance achieved are not guaranteed to be portable to other graph databases. We intend to address this limitation in our future work. Other directions for future work are to evaluate these encoding methods on more compliance patterns and more event logs. Furthermore, we intend to translate several analysis queries against graph-encoded event logs available in the literature into queries against these encoding schemes to assess whether there is still a performance gain and graph size reduction.

\small

\bibliographystyle{splncs04}
\bibliography{bibliography}

\end{document}